\documentclass{sigchi}

\toappear{\scriptsize This is the pre-print for a poster paper accepted to UIST '20\\ }
\usepackage{balance}       % to better equalize the last page
\usepackage{graphics}      % for EPS, load graphicx instead 
\usepackage[T1]{fontenc}   % for umlauts and other diaeresis
\usepackage{txfonts}
\usepackage{mathptmx}
\usepackage[pdflang={en-US},pdftex]{hyperref}
\usepackage{color}
\usepackage{booktabs}
\usepackage{textcomp}

\usepackage{microtype}        % Improved Tracking and Kerning
\usepackage{ccicons}          % Cite your images correctly!
\def\plaintitle{Adapting Nielsen's Usability Heuristics to the Context of Mobile Augmented Reality}

\def\emptyauthor{}
\def\plainkeywords{augmented reality, usability heuristics}

\makeatletter
\def\url@leostyle{%
  \@ifundefined{selectfont}{
    \def\UrlFont{\sf}
  }{
    \def\UrlFont{\small\bf\ttfamily}
  }}
\makeatother
\def\pprw{8.5in}
\def\pprh{11in}

\definecolor{linkColor}{RGB}{6,125,233}
\hypersetup{%
  pdftitle={\plaintitle},
  pdfauthor={\emptyauthor},
  pdfkeywords={\plainkeywords},
  pdfdisplaydoctitle=true, % For Accessibility
  bookmarksnumbered,
  pdfstartview={FitH},
  colorlinks,
  citecolor=black,
  filecolor=black,
  linkcolor=black,
  urlcolor=linkColor,
  breaklinks=true,
  hypertexnames=false
}

\def\plaintitle{Adapting Usability Heuristics to the Context of Mobile Augmented Reality}

\def\emptyauthor{}
\def\plainkeywords{augmented reality, usability heuristics}

\begin{document}

\title{\plaintitle}

\numberofauthors{2}
% Notice how author names are alternately typesetted to appear ordered
% in 2-column format; i.e., the first 4 autors on the first column and
% the other 4 auhors on the second column. Actually, it's up to you to
% strictly adhere to this author notation.
\author{%
    \alignauthor{%
    \textbf{Audrey Labrie}\\
    \affaddr{Polytechnique Montr{\'e}al} \\
    \affaddr{Montr{\'e}al, Qu\'ebec, Canada} \\
    \email{audrey.labrie@polymtl.ca} }
    \alignauthor{% 2nd
    \textbf{Jinghui Cheng}\\
    \affaddr{Polytechnique Montr{\'e}al} \\
    \affaddr{Montr{\'e}al, Qu\'ebec, Canada} \\
    \email{jinghui.cheng@polymtl.ca}}
}

\maketitle

% Do not change the page size or page settings.
\begin{abstract}
Augmented reality (AR) is an emerging technology in mobile app design during recent years. However, usability challenges in these apps are prominent. There are currently no established guidelines for designing and evaluating interactions in AR as there are in traditional user interfaces. In this work, we aimed to examine the usability of current mobile AR applications and interpreting classic usability heuristics in the context of mobile AR. Particularly, we focused on AR home design apps because of their popularity and ability to incorporate important mobile AR interaction schemas. Our findings indicated that it is important for the designers to consider the unfamiliarity of AR technology to the vast users and to take technological limitations into consideration when designing mobile AR apps. Our work serves as a first step for establishing more general heuristics and guidelines for mobile AR.
\end{abstract}

% ACM Classfication
% \begin{CCSXML}
% <ccs2012>
%   <concept>
%       <concept_id>10003120.10003121.10003124.10010392</concept_id>
%       <concept_desc>Human-centered computing~Mixed / augmented reality</concept_desc>
%       <concept_significance>500</concept_significance>
%       </concept>
%   <concept>
%         <concept_id>10011007.10010940.10011003.10011687</concept_id>
%         <concept_desc>Software and its engineering~Software usability</concept_desc>
%         <concept_significance>500</concept_significance>
%         </concept>
%  </ccs2012>
% \end{CCSXML}

% \ccsdesc[500]{Human-centered computing~Mixed / augmented reality}
% \ccsdesc[500]{Software and its engineering~Software usability}

% Print the classficiation codes
%\printccsdesc

%\keywords{\plainkeywords}

\section{Introduction}
Mobile augmented reality (AR) is gaining an increasing attention and started to break into the mass market due to the availability of the technology on major mobile platforms (e.g. iOS and Android). %Applications leveraging this technology have experienced a rapid growth and started to break into the mass market. 
Although an increasing number of users are discovering mobile AR, there are no commonly accepted guidelines or heuristics to support the designers of this type of applications to create an optimized user experience~\cite{dunser2007applying}. We aim to address this challenge by identifying these guidelines and heuristics from the evaluation of the existing commercial-off-the-shelf applications. As a first step towards this goal, we examine how Jakob Nielsen's Usability Heuristics~\cite{Nielsen1990}, a set of widely-known heuristics for general applications, can be applied to the context of mobile AR design.

Particularly, we conducted a case study on AR home design apps. We chose this application domain because, from our preliminary investigation of mobile AR applications, home design is a domain where a lot of commercial-off-the-shelf apps are available and are created by professional companies. Additionally, this type of applications has common patterns of interactions that are important in most mobile AR applications (i.e. placing and manipulating virtual objects in a scene)~\cite{Sharma2018}. Finally, the design and evaluation of AR home design apps is an active research direction. For example, Park et al.~\cite{Park2017, Park2016} developed a questionnaire, based on usability analysis guidelines~\cite{Atkinson2007} and the usability principles~\cite{dunser2007applying}, for evaluating an AR home design application. Viyanon et al.~\cite{Viyanon2017} also created a questionnaire based on Nielsen's definition of usability. 

This paper situates in these related studies and focuses on interpreting and adapting Nielsen's heuristics in the AR home design context. Through this effort, we provide a first step towards the interpretation of Nielsen's usability heuristics for mobile AR applications, as well as the creation of specialized guidelines and heuristics for mobile AR application design.

\section{Methods}
To select a set of apps for analysis, we first used the search string "augmented reality home design" in the Apple App Store and found 14 apps. To narrow down our selection, we excluded those that did not update within a month to focus on the active projects. Additionally, we added one other app (i.e. Sayduck) because it had unique features. The apps selected this way were: IKEA Place, Houzz Home Design Renovation, Interior Define AR, Stressless\@home, Graham Brown Design Renovation, and Sayduck. The common goal of the AR home design apps is to enable the user to view virtual objects (e.g. furniture, wallpapers, and tiles) in a physical environment to get a sense of how it would look in real life.

After selection of the mobile applications, we analyzed each apps by answering thoroughly the following questions: (Q1) How can users interact with the AR objects? (Q2) What kind of information is presented on the mobile screen and how? (Q3) What aspects of the app are well designed? (Q4) What are the prominent usability issues? The two authors first independently reviewed the two most popular apps (IKEA Place and Houzz Home Design) and discussed their findings. Then, we collaboratively mapped each feature and issue identified in those questions to the Nielsen's 10 Usability Heuristics. This served as a groundwork to the first draft of interpretation of Nielsen's heuristics in AR home design apps. The other apps were reviewed using the same process and the interpretation of the heuristics were modified iteratively.

\section{Results}
The following interpretation of Nielsen's heuristics is based on both the usability issues and the well-designed features identified in the analyzed AR home design apps.

\textit{Visibility of system status.}
The system should keep the user informed during the interaction. When scanning surfaces, for example, the user should be informed when the physical surfaces are difficult to detect (e.g., too reflective or too dark). After successful surface detection, the system should notify the user that the object can be placed. If it takes time to load an object, the system should inform the user.

\textit{Match between system and the real world.}
Virtual objects should appear and behave as realistically as possible. Their sizes should be proportional to the physical environment and they should be fixed to surfaces, as it is in the real world. Objects should not overlap with each other, i.e., collisions between objects should be detected by the system.

\textit{User control and freedom.}
It is possible that the user places a virtual object in an unwanted location, or selects, deletes or rotates the object by mistake. Therefore, the system should support undo and redo and it should confirm with the user when deletion is selected.

\textit{Consistency and standards.}
Gestures used to translate and rotate an object should be intuitive. It is particularly important to indicate to the user how the object can be manipulated. If adjusting the height is required for an object, it should be done by an extra UI component (e.g. a vertical slider) because directly moving the object on the vertical dimension can be easily confused with object translation.

\textit{Error prevention.}
Possible errors include placing or deletion of an object by mistake. For these error-prone actions that can lead to consequences and inconveniences, the system should ask for confirmation. If the user does not confirm, it would return the object to its original location. %For deletion, the system should explicitly ask the user for confirmation when deletion is selected.
% This heuristic is also related to using intuitive gestures for the manipulation of objects (\textit{Consistency and standards}). This way, it would help prevent the user from placing an object at an unwanted location or orientation. 

\textit{Recognition rather than recall.}
Possible actions regarding interaction with virtual objects should be always visible. When adding an object to the scene, to avoid recall (i.e. remembering what object is to be added), the system should display an image of the selected object. When an object is selected, all possible actions should be shown on the screen (e.g. undo/redo, delete, change color); useful gestures for translating and rotating the object should also be indicated. When no object is selected, the system should show the possible actions related to the entire scene (e.g. take a photo, favorites, reset scene).% and an option to go back to the main menu.

\textit{Flexibility and efficiency of use.}
%Flexibility and efficiency of use include features tailored to frequent actions and accelerators. 
Regarding frequent actions, the system should allow the user to place several objects on the same scene at the same time. %For example, when designing a living room, the user could select all relevant objects and then navigate to the scene to place them all at once. 
Regarding accelerator, the presence of additional information about objects such as measurements, physical front, material, and price would be helpful. If it is technologically possible, objects should be automatically snapped to a physical corner, which would reduce the number of actions needed by the user.

\textit{Aesthetic and minimalist design.}
The system should not show irrelevant or rarely used information to the user. Extra information could be shown when the user makes the corresponding selection. Examples of the common relevant information during the different stages of interaction are: (1) when placing an object, there should be a placement shape indicating the position and the size of the object; and (2) when selecting an object, an indicator of selection and options for object operations should be displayed.

\textit{Help users recognize, diagnose, and recover from errors.}
Help for error recovery is important. %When errors with AR functionalities occur, the system should inform the user of how to recover from those errors. 
For example, the moment the system has trouble detecting surfaces, the user should be informed of the reason and help the user correct the error. %For example, the system should let the user know explicitly if the surface is too reflective or too dark to be detected.
If the user is moving a virtual object on an undetected surface, the user should have the option to re-activate surface detection. %When detected, the object should be moved on the corresponding surface.

\textit{Help and documentation.}
Such a system usually should not have explicit documentation because it should be intuitive to use. However, they should have a brief tutorial for first time users. %What it should have is help at different stages of interaction because users can be unfamiliar with AR apps. 
Particularly, surface detection is an action that most users are not familiar with. So there should be a brief tutorial and explicit indication telling the user how to detect surfaces. The possibility of translating and rotating the objects should also be clearly communicated to the user.
\section{Discussion}
%In this section, we discuss several reflection points based on our analysis of the current AR home design applications and the mapping of their usability issues with Nielsen's heuristics.

One of our prominent findings is that many usability issues in AR home design apps are associated with the current limitations of AR technology. For example, one of the most problematic aspects of these apps is the incorrectness or inability to detect surfaces. %Not all surfaces can be detected, some are too dark or too reflective. 
%Because of this, in all apps, virtual furniture could be placed in physically impossible locations (e.g. a chair can float in the air). 
These technological limitations can affect usability. For example, if a surface is too dark to be detected and the system does not tell the user, the user will only be left wondering why surface detection is taking a long time and not working. Therefore, when developing an AR mobile app, identifying technological limitations and taking them into consideration during the design process is critical.

Further, because mobile AR is still an emerging technology, most AR app users should be considered as novices. For these users, several actions such as surface detection or object manipulation (i.e. interactions with virtual 3D objects on a 2D space) can be confusing. Therefore, it is important to show the users how to use the technology and how to interact with the objects with gestures. One solution could be to show animation gestures during the first time of use. Also, confusing interactions need to be avoided. For example, there should be two different types of interactions for adjusting the height of an object and vertical translation (moves the object closer or further from the user) because intuitively, the gesture for both actions would be swiping up and down. %To make the interaction as realistic as possible, translation should move the object fixed to a surface and if the height has to be adjusted (e.g. the object is attached to the ceiling), a vertical slider should be used.

As previously mentioned, the mapping to Nielsen's heuristics is specific to AR mobile home design apps and they could be used to establish more general heuristics and guidelines about mobile AR design. %The purpose of this study was not to create new heuristics but to interpret a set of classical heuristics in a new design context. 
To evaluate their validity, more analysis on other applications is needed and a user study where UX designers use the heuristics to design or evaluate AR home design interfaces could be done in future work.

% Most surprising results/why are they surprising?
% - might not be adapted for novice users
% - technological limitation impact on usability
% - 3D objects well made but gap with intuitive interactions and overall experience with the apps
% - all apps had (different) issues

% Main findings about the current landscape of AR apps
% How the interpretations of the heuristics can be helpful - for design and evaluation of AR apps
% - From these heuristics, since they are based on both positive and negative elements of current AR apps, they could be consulted at the design stage. It could help avoid common mistakes and build based on good practices.
% - Evaluation of AR apps: 

\section{Conclusion}
In this study, we focused on interpreting Nielsen's 10 Usability Heuristics in the context of AR home design mobile applications. To do this, we evaluated six AR home design apps available on the Apple App Store to identify their well-designed features and usability issues to map them to Nielsen's heuristics. %From this, we adapted the heuristics to establish what each meant in the context of mobile AR for home design. 
This effort serves as a first step in establishing more full-fledged heuristics and guidelines for the design of mobile AR apps.

\balance{}

\newpage
\bibliographystyle{SIGCHI-Reference-Format}
\bibliography{references}
\end{document}